\documentclass[10pt, conference, compsocconf]{IEEEtran}
\usepackage{epsfig}
\usepackage[T1]{fontenc}
\usepackage{amssymb}
\usepackage{amsmath}
\usepackage{amsfonts}
\usepackage{verbatim}
\usepackage{graphicx}
\usepackage{hyperref}
\usepackage{array}
\usepackage{multirow}
\usepackage{epstopdf}
\usepackage{color}  
\begin{document}

\title{FPGA based Novel High Speed DAQ System Design with Error Correction}
\author{\IEEEauthorblockN{Swagata Mandal\IEEEauthorrefmark{1},
Suman Sau \IEEEauthorrefmark{2},
Amlan Chakrabarti \IEEEauthorrefmark{2}, 
Jogendra Saini \IEEEauthorrefmark{1}, \\
Sushanta Kumar Pal \IEEEauthorrefmark{1} and
Subhasish Chattopadhyay \IEEEauthorrefmark{1}}
\IEEEauthorblockA{\IEEEauthorrefmark{1}Variable Energy Cyclotron Centre, 
Kolkata, India \\ \IEEEauthorrefmark{1}{(swagata.mandal, sushant, sub)}@vecc.gov.in}
\IEEEauthorblockA{\IEEEauthorrefmark{2}A.K.Choudhury School of Information Technology, University of Calcutta, Kolkata, India\\
\IEEEauthorrefmark{2}{(ssakc\_s, acakcs)}@caluniv.ac.in}

}

\maketitle
\begin{abstract}

Present state of the art applications in the area of high energy physics experiments (HEP), radar communication, satellite communication and bio medical instrumentation require fault resilient data acquisition (DAQ) system with the data rate in the order of Gbps. In order to keep the high speed DAQ system functional in such radiation environment where direct intervention of human is not possible, a robust and error free communication system is necessary. In this work we present an efficient DAQ design and its implementation on field programmable gate array (FPGA). The proposed DAQ system supports high speed data communication ($ \mathtt{\sim}$4.8 Gbps) and achieves multi-bit error correction capabilities. BCH code (named after Raj Bose and D. K. RayChaudhuri) has been used for multi-bit error correction. The design has been implemented on Xilinx Kintex-7 board and is tested for board to board communication as well as for board to PC using PCIe (Peripheral Component Interconnect express) interface. To the best of our knowledge, the proposed FPGA based high speed DAQ system utilizing optical link and multi-bit error resiliency can be considered first of its kind. Performance estimation of the implemented DAQ system is done based on resource utilization, critical path delay, efficiency and bit error rate (BER).
\end{abstract}

%A category including the fourth, optional field follows...
%\category{D.2.8}{Software Engineering}{Metrics}[complexity measures, performance measures]
%\keywords{DAQ, FPGA, SEU, Error correction, PCIe, SGDMA} % NOT required for Proceedings
\section{Introduction}
\label{Introduction}
 High speed and fault resilient DAQ system is an integral part of the signal processing unit in some crucial real time applications like radar communication, HEP, satellite communication etc. In a traditional DAQ system Frontend Electronics (FEE) board captures data from the sensors through high speed LVDS link, processes it and sends it to storage device using high speed link like Ethernet, PCIe, fiber optic etc. for further analysis. Commonly faced problems in the traditional DAQ systems are low data rate~\cite{Wang:iciea:2009} and prone to SEU in highly radiated area. Presently optical fiber and PCIe are the most suitable options in high speed data transmission over normal copper based LVDS line as they are susceptible to noise and interference.
 \par 
 In~\cite{Minami:ieeetran:2011} the authors developed a new Gigabit Optical Serial Interface Protocol (GOSIP) for communication over optical fiber  and implemented PCIe to optical link interface on FPGA for their DAQ system. This system gives a stable data rate of 1.6 Gbps. In~\cite{kadric:socc:2012}\iffalse, we find the \fi  development of an optical link between two PCIe buses of computing nodes with data rate of 8.5 Gbit/s has been proposed. Authors used PCIe hard IP available in ALTERA Stratix IV FPGA board. Liansheng Liu \textit{ et.al} presented the development of a fibre channel node with PCIe interface for avionics environments in~\cite{Liu:i2mtc:2013:X}. Each node consists of two modules: FPGA module and PowerPC module. Two nodes are connected by optical fiber through Small Form-factor Pluggable (SFP) interface and maximum data transfer rate achieved in this case was 2.125 Gbps. A high-speed data transmission protocol over optical fiber for real time data acquisition in Beijing Spectrometer III (BESIII) trigger system had been developed by Hao Xu \textit{et.al} in~\cite{haoxu:nss:2007:XX}. Here they have used Multi Gigabit Transceiver (MGT) of Virtex-II Pro series FPGA for data transmission over optical fiber and achieved data rate of 1.75 Gbps. In~\cite {mattihalli:cecnet:2012}, a high speed data transfer protocol named Serial Front Panel Data Port (SFPDP) over optical fiber is implemented on FPGA to capture the data from Digital Signal processor (DSP) through Extended Attachment Unit Interface (XAUI) where the optical link can work on three distinct speeds: 1.0625 Gbaud, 2.125 Gbaud, and 2.5 Gbaud. In~\cite{bohm:nss:mic:2012}, authors have used a bus master DMA along with a 4-lane generation 2 PCIe link to transfer the stream data from FPGA to PC. The highest data transfer rate achieved in this case is 784 Mbps.
\par
Normally, SEU occurs when a charged particle hits and transfers sufficient energy to the silicon area of a circuit.  
The SEU mitigation techniques can be classified into two types: prevention and recovery. Prevention methods are mainly considered during the ASIC design. The recovery methods include online recovery mechanisms \textit{e.g.} fault tolerant computing, error detecting/correcting code and online testing, which make the system more robust. Several error detection/correction techniques have been tried by many researchers. The concurrent error detection (CED)~\cite{Siewiorek:CED} is one of such technique where an extra error detection circuit is used along with the main circuit. When an error is detected, the main circuit recomputes or rolls back the whole operation from the beginning. Triple Modular Redundancy (TMR)~\cite{TMR:IeeeTran:NuclrPhy} is another scheme where the same functional replica is used thrice and the final result is based on majority voting system. Above mentioned schemes are not applicable in real time because they are based on either time or space redundancy or combination of both. The error detection and correction (EDAC) codes play an important role in many successful SEU mitigation schemes. The SEU can be protected by using single error correcting hamming code, which may not be sufficient for reliable communication in high speed DAQ system. So multiple error correction codes can be applied in this kind of high speed DAQ. Several multiple forward error correction codes (FEC) have been presented in different papers \textit{e.g.} BCH code~\cite{chien:search}, Reed Solomon code and Reed Muller code~\cite{ Varghese:multibit:error}. %\cite{hanho:ieeetran:2005:5}
\par
In all of the above mentioned works the authors did not discuss anything about SEU mitigation in high speed data acquisition system to work in an adverse environmental condition like Deep Space Experiment or in HEP experiment. The scope of our work takes into consideration an efficient design of the various stages in the DAQ chain to meet the high data  rate requirements. Our high speed data acquisition system is protected from SEU by multi-bit error correcting code and interleaver. Scrambling is used here as line coding technique to maintain the DC balance and to obtain 20\% extra throughput unlike 8b/10b coding in~\cite{Minami:ieeetran:2011},~\cite{kadric:socc:2012},~\cite{Liu:i2mtc:2013:X}. We have achieved a maximum data rate of 4.8 Gbps compared to 1.6 Gbps,2.125 Gbps,1.75 Gbps in~\cite{Minami:ieeetran:2011},~\cite{Liu:i2mtc:2013:X},~\cite{haoxu:nss:2007:XX} respectively.
In this paper, our key contributions are:
\begin{itemize}
\item Efficient implementation of FPGA based high speed DAQ with optical link having multi-bit error correction capability.
\item Performance analysis of the DAQ has been done using real time test set up having PCIe gen2 and scatter gather direct memory access (SGDMA).
\end{itemize}   
The rest of the paper is organized as follows. Section~\ref{SystemDesignDAQ} describes the full system design topology for the high speed DAQ. Experimental setup with performance evaluation are described in Section~\ref{PerformenceEvoluation} followed by concluding remarks in Section~\ref{Conclusion}.

\section{System design for high speed daq}\label{SystemDesignDAQ}
\begin{figure}[!htb]	
%\centering
\hspace{-20 pt}
\vspace{-5pt}
\includegraphics[scale=0.21]{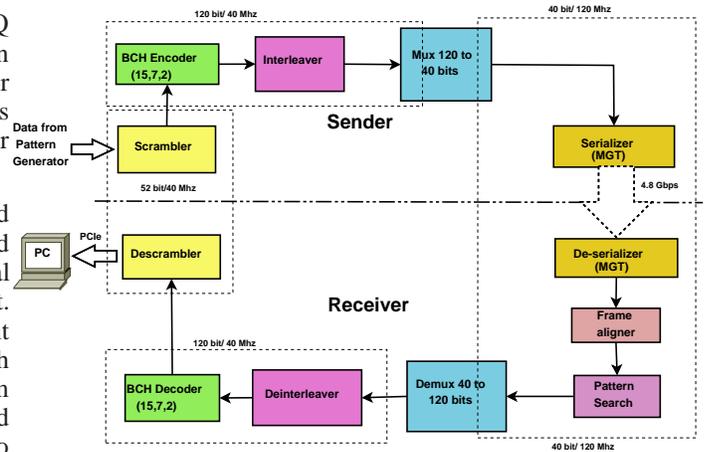}
\caption{Internal blocks of the proposed system}
\vspace{-15pt}
\label{fig:BlockDiagramFlow}
\end{figure}

The main aim of the DAQ used in critical applications are to handle high data rate, error correction capabilities and efficient storage mechanism for future analysis. In high speed data transmission optical fiber is generally used as the communication media. Several multibit error correction methods for efficient communication had been discussed in Section~\ref{Introduction}, where BCH coding is most suitable for random error correction. The interleaver block has been introduced after encoder block judiciously to enhance the error correction efficiency. In the receiver side data is directly transfered to PC through PCIe from the FPGA board. Functional blocks of the proposed system is shown in Figure~\ref{fig:BlockDiagramFlow}. The details of each block have been discussed in the following subsections. 
%----------------------
\subsection{Scrambler/Descrambler} Scrambler is used to reduce the occurrence of long sequences of `1' (or `0') that maintains a good DC balance in input signal coming from the detector/sensor. This helps in accurate timing recovery on receiver equipment. It has a latency of one clock cycle but does not add any overhead in the system like the $8b/10b$ or $7b/8b$ line coding. Here 52 bit incoming data is divided into four blocks of 13 bit data and then each block is scrambled simultaneously using 13 bit polynomial. These scrambled data from all four blocks are combined together to produce the 52 bit scrambled output data. 
\subsection{BCH Encoder/Decoder} BCH is a binary error correcting code. Here, BCH (15,7,2) code is used to correct the error due to SEU or Multiple bit upset (MBU). In this coding scheme 7 bit data is appended with 8 redundant bits for error correction. So the code rate (ratio of input data to coded data) is 0.467. 56 bit data (52 bit data with 4 bit header) is broken down into eight 7 bit data, which are encoded with BCH encoder in parallel. After encoding the 15 bit data from the eight blocks, they are assembled to generate a total 120 bit data. Though this block introduces one clock cycle latency in the system but increases the reliability in data transfer. The encoded data has been decoded in the following three steps: determination of the error locater polynomial, detection of error location using Chien Search Algorithm~\cite{chien:search} and location of the data at the error position\iffalse One can find the details of BCH algorithm in~\cite{bch:book} \fi. In our present work we have designed the BCH encoding/decoding block as a custom hardware design. Instead of selecting BCH code with larger block size like (31, 26, 1) or (63, 57,1), we used eight BCH (15,7,2) in parallel for faster error correction without compromising the time complexity. Hence, each BCH decoder block can correct up to 2 bits of error within 7 bits of input. So the total  $8\times 2 = 16$  error can be corrected simultaneously using this technique without any extra resources. Similarly, we can use triple error correcting BCH code~\cite{chien:search} but that will reduce the code rate.
%\begin{enumerate}
% \item Determination of the error locater polynomial
%      \item Detection of error location using Chien %Search Algorithm~\cite{chien:search}
 %           \item Location of the data at the error %position
%      \end{enumerate} 
\subsection{Interleaver/De-interleaver} Interleaving is the reordering of the data that is to be transmitted, so that the consecutive bytes of data are distributed over a larger sequence of data to reduce the effect of burst error. Generally two types of interleaving strategies  (Block interleaver and convolutional interleaver) are used in any communication system. Here we have used block interleaver. The first 120 bit data from the encoder is divided into two block of 60 bits of data and then interleaving operation is done on each 60 bit data using block interleaver. The whole process increases the code correction capabilities without any clock latency and overhead. De-interleaver process is used to reorder the data again in the receiver side.
\subsection{MUX/DEMUX and Clock Domain Crossing} 
\begin{figure}[htb]
\hspace{-30 pt}
\includegraphics[scale=0.25]{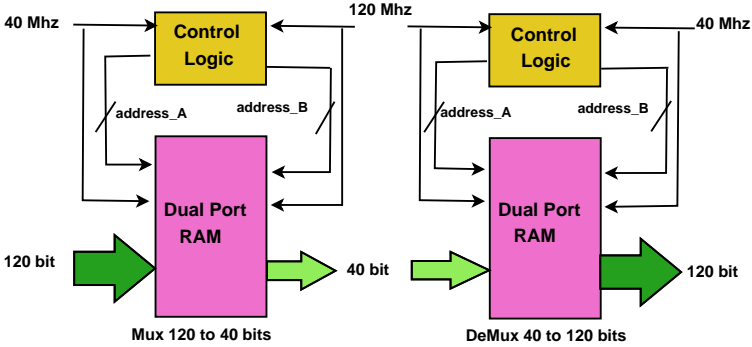}
\vspace{-14 pt}
\caption{Mux DeMux for clock domain crossing}

\label{fig:MuxDMux}
\end{figure}

This block consists of dual port RAM and read-write controller. It breaks down 120 bit frame into three words of 40 bits width. It reduces bandwidth consumption keeping data rate same. Here, we have used 120 MHz clock to drive the multi-gigabit transceiver (MGT) available from Xilinx IP core to keep the data rate same with the internal blocks those are running with 40 MHz frequency. The data rate and clock frequency can be changed to any value according to the requirement. This block is used to synchronize the data rate between MGT and the other parts of the design. Figure~\ref{fig:MuxDMux} shows the architectural block diagram of the MUX-DEMUX and clock domain crossing.
\subsection{Serializer/De-serializer} This block simply converts the parallel data to serial data, which is transmitted over the communication channel. It is inbuilt within the MGT.  De-serializer simply converts the serial data to parallel data in the receiver side.
\subsection{Frame Aligner and Pattern Search} \label{FrameAligner}
\begin{figure}[t!]
\centering
\includegraphics[scale=0.30]{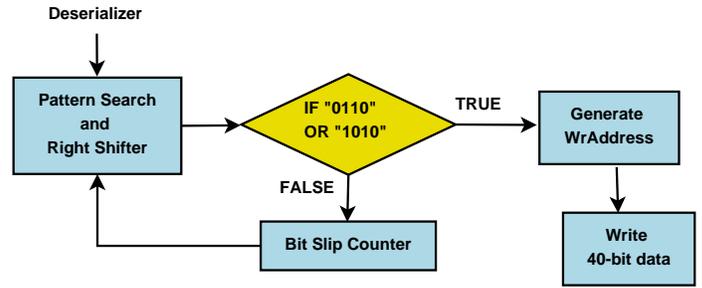}
\vspace*{-10pt}
\caption{Algorithm for Frame Aligner and Pattern Search}
\vspace*{-5pt}
\label{fig:FrameAlignerFlow}
\end{figure}

\begin{figure}[t!]
\centering
\includegraphics[scale=0.31]{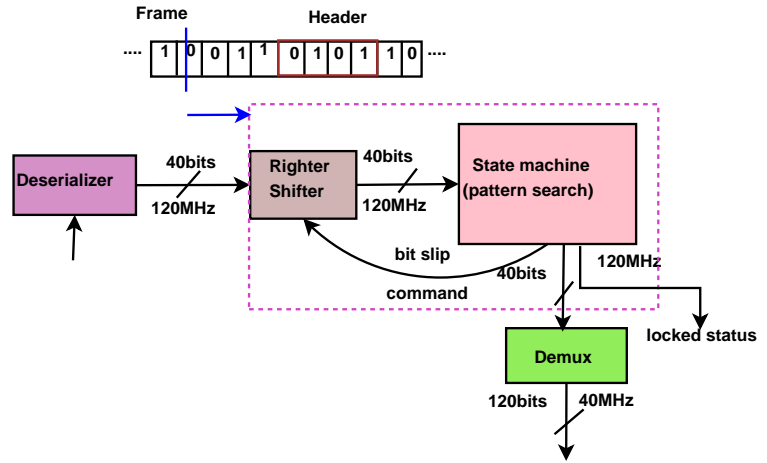}
\vspace{-5 pt}
\caption{Data flow diagrams of the Frame Aligner and Pattern Search block}
\vspace{-10 pt}
\label{fig:FrameAlignerWork}
\end{figure}

In the receiver side the frame aligner block aligns the frames in a proper order by using frame header as an index. This frame header is detected by an efficient pattern search algorithm whose flow chart is given in Figure~\ref{fig:FrameAlignerFlow}. There are two types of frame: Standard and Frame without FEC.% %----------------------
The standard frame consists of four fields: Header field (4 bit), Slow Control (4 bit), Data field of width 48 bit, FEC field of width 64 bit. 
Slow Control (here, we merged with data field) field is reserved for controlling the DAQ chain in future.
Whereas frame format without FEC consists of three field: Header field (4 bit), Slow control field (4 bit) and data field (112 bit). In standard frame format $1010$ is used as header and $0101$ is used as header in frame format without FEC. Frame format without FEC may be used for those applications where probability of error is low and high throughput is needed. The frame aligner and pattern search block consists of two sub blocks (Pattern search block, Right shifter block) as shown in the Figure~\ref{fig:FrameAlignerWork}.
  Right shifter block shifts the received data by one bit to the right from MSB side and sends it to the pattern search block. The pattern search block checks whether the header is received or not. Once the header is properly detected pattern search block will continuously search for another 32 subsequent headers of other frames to ensure the stability of the link and then the process gets terminated.
%%----------------------
%\begin{figure}[h]
%\centering
%\includegraphics[scale=0.25]{Wr40BitAddrFlow.eps}
%\caption{Algorithm for Frame Aligner and Pattern Search}
%\vspace*{-8pt}
%\label{fig:FrameAlignerFlow}
%\end{figure}

%------------------------

\subsection{Data Transfer to Host PC through PCIe}
\begin{figure*}[htb]
\centering
\includegraphics[scale=0.24]{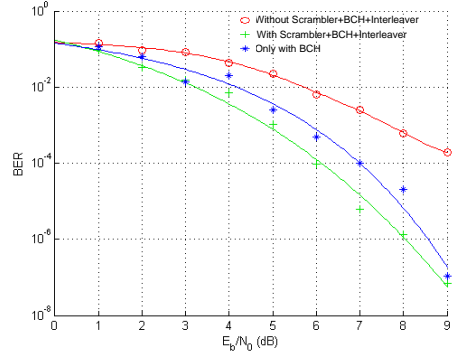}
\vspace{-5 pt}
\caption{ PCIe interfacing with blocks and Experimental setup of proposed DAQ}
\label{fig:PCIeSetup}

\end{figure*}

The asynchronous Fast In Fast Out (FIFO) and SGDMA is used to transfer data from FPGA board to PC through PCIe. We have used PCIe gen2 Intellectual Property (IP) core available from Xilinx. Interconnection of FPGA to PC through PCIe is shown in Figure~\ref{fig:PCIeSetup}. Data is written into FIFO at a frequency of 120 MHz by which MGT is running and  data will be read from FIFO at a frequency of 125 MHz by which PCIe core is running. In the PC side we capture the data by a program developed using windows software development kit (SDK) written in C language.  

%\section{System Implementation} \label{SystemImplementation}
\subsection{Data Flow Overview}
\begin{figure*}[htb]
\centering
\includegraphics[scale=0.20]{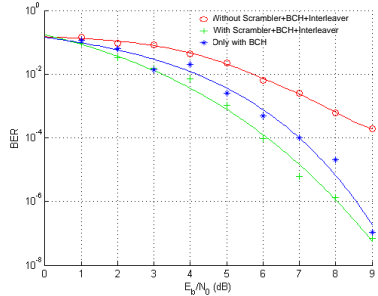}
\vspace{-10 pt}
%\hspace{-20 pt}
\caption{Standard frame generation and Error correction flow}
\label{fig:data_flow}
\vspace{-15 pt}
\end{figure*}

 The complete chain of the functional blocks as shown in Figure~\ref{fig:BlockDiagramFlow} for the high speed DAQ with multi-bit error correction (Considering two bits error correction) has been implemented on the FPGA board. Figure~\ref{fig:data_flow} shows the complete mechanism of standard frame generation and the error correction flow.
 At first, only 52 bit user data is scrambled by the scrambler block. These 52 bits of data is divided into four blocks of 13 bits data and scrambles each block parallely. The scrambled data with the 4 bit header is mapped in the input lines of the eight BCH (15,7,2) encoders, which are running parallel.  Output of all the encoders are combined to get a frame of 120 bit data. This 120 bits of data are interleaved first and then goes to the next functional block that is the MUX. Interleaving is used to reduce the effect of burst error. But the header position is never changed in the frame format (red color in Figure~\ref{fig:data_flow}) even after interleaving process, helps to synchronize the frame in the receiver side. In Mux-DeMux and clock domain crossing block a dual port RAM is used to write this 120 bits data using 40 MHz clock and read the same data at 120 MHz clock rate with 40 bit word size. So the data rate for writing ($40 \times 120=4.8 $ Gbps) and reading ($120\times 40=4.8$ Gbps) are same. The 40 bit data is serialized first and goes to the transmitter (TX) for transmitting over the optical fiber cable. In the receiver (RX) side functional blocks are Deserializer, DEMUX, De-interleaver, BCH Decoder (15, 7, 2) and Descrambler. They perform reverse function with respect to Serializer, MUX, Interleaver, BCH Encoder (15, 7, 2) and Scrambler respectively. The extra block frame aligner and pattern search in the receiver side is added in this chain whose functional description has been described in section~\ref{FrameAligner}.

%%-----------------------------
%\begin{figure*}[h]
%\centering
%\includegraphics[scale=0.30]{pcilayer.eps}
%\vspace*{-8pt}
%\caption{Structure of PCIe layers}
%\label{fig:pcilayer}
%\end{figure*}
% %----------------------

\iffalse.Figure~\ref{fig:FrameAlignerWork} shows how this block works.\fi   
%----------------------
%------------------------
%\section{System Implementation} \label{SystemImplementation}
\section{Experimental Setup and performance Analysis} \label{PerformenceEvoluation}
\begin{figure*}[htb]
%\hspace{-10 pt}
\includegraphics[scale=0.45]{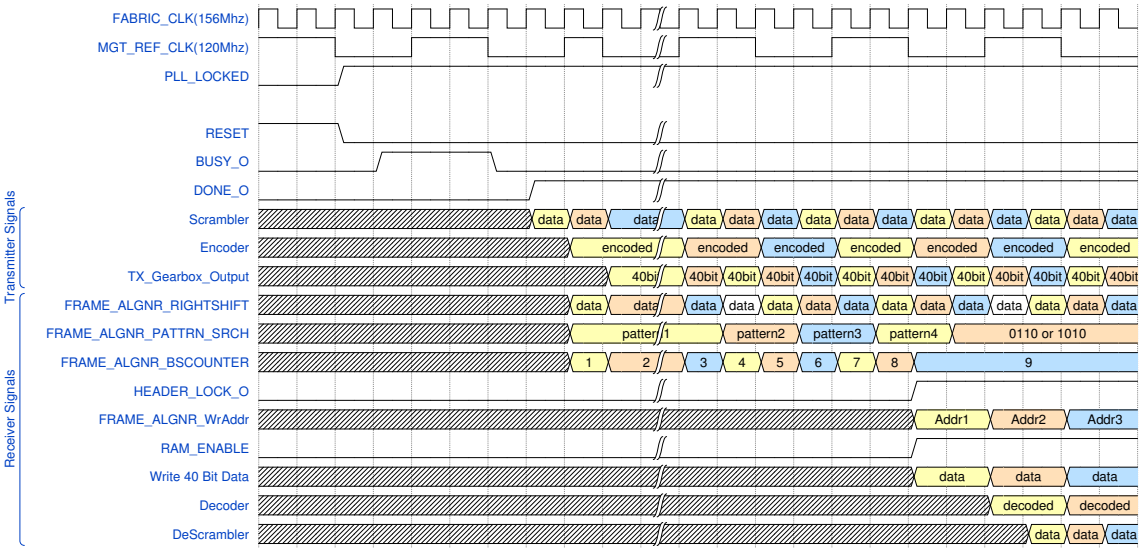}
%\vspace{-10 pt}
\caption{Timing diagram of the transmitter and receiver signals }
%\vspace*{-8pt}
\label{fig:GBTTimingTx}
\end{figure*}

The full prototype of the DAQ chain is implemented in the Xilinx Kintex-7 boards (KC705 from Avnet) using the Xilinx ISE 14.5 platform and VHDL for design entry. We have used an external jitter cleaned clock source (CDCE62005EVM of TI) to drive MGT of two Kintex-7 boards. One Agilent power supply has been used to drive the whole system. Two Kintex-7 boards are connected through single mode optical fiber using SFP from \textit{Finisar} (FTLX8571D3BCL). For board to PC communication we have used eight lane PCIe gen2. Names of the various signals and their functions are given in Table~\ref{table:SignalName}.
The timing diagram of the transmitter and the receiver side are given in Figure~\ref{fig:GBTTimingTx}. 
%================================================================
\begin{table*}[htb]
\begin{center}
\caption{Description of the signals used in timing diagram}
\scalebox{1.1}{
\begin{tabular}  {|c|c|c|c|} %{ | m{5em} | m{2cm}| m{2cm} | } % 
\hline
\textbf {Signal} & \textbf{Width} & \textbf{Function} & \textbf{Use} \\ 
\hline
Fabric\_Clk & 1 & Use to drive different blocks in DAQ & Both in Transmission and Receiver side \\ 
\hline
MGTREF\_Clk  &1 & Use to drive MGT & Both in Transmission and Receiver side \\ 
\hline
PLL\_Clk & 1 & Use to drive MGT & Both in Transmission and Receiver side \\ 
\hline
PLL\_Locked  &1 & Output of PLL. It indicates PLL generate stable clock & Both in Transmission and Receiver side\\
\hline
RESET  &1 & Use to reset the whole system & Both in transmission and Receiver side\\
\hline
BUSY\_O  &1 & Becomes high when System enters a process before ready & Both in transmission and Receiver side\\
\hline
Scrambler  & 52& This signal contains the data of output of scrambler block & In Transmission side only\\
\hline
Encoder  & 120& This signal contains the data after BCH encoding & In Transmission side only\\
\hline
MUX\_Output  &40 & This signal contains output of MUX block which is 40 bit width & In Transmission side only\\
\hline
FRAME\_ALIGNR &&& \\

\_RIGHTSHIFT  &40 & Store the receive data after shifting right by one bit & In Receiver side only\\
\hline
FRAME\_ALIGNR &&&
\\
\_PATTERN\_SEARCH  &4 & Check whether header is matched or not & In Receiver side only\\
\hline 
FRAME\_ALIGNR &&&
\\
\_BSCOUNTER  & 5& Store the output of counter until header is not matched & In Receiver side only\\
\hline 
Header\_LOCK\_O  & 1& Becomes high when the frame is locked & In Receiver side only\\
\hline
FRAME\_ALIGNR &&&
\\
\_WrAddr  & 5 & store the address of RAM where receive data will be written & In Receiver side only\\
\hline
RAM\_ENABLE  &1 & Becomes high when RAM is Ready to perform & In Receiver side only\\
\hline
Write 40 bit Data  &40 & Store 40 bit data which is to be written in RAM &  In Receiver side only\\
\hline
DECODER  & 52& Contains the decoded data & In Receiver side only\\
\hline
DESCRAMBLER  & 52& Contains output data of descrambler block & In Receiver side only\\  
\hline
\end{tabular}}
%\caption{Description of the signals used in timing diagram}
\vspace{-10 pt}
\label{table:SignalName}
\end{center}
\end{table*}
%\section{Experimental Setup and performance} \label{PerformenceEvoluation}
The block diagram and experimental setup of the system are shown Figure~\ref{fig:PCIeSetup}. We achieved maximum bit rate of 4.8 Gbps in our system. In standard mode, a frame contains only 52 bits of data, 64 bits for error correction (FEC) and 4 bits of header. 64 bits  FEC field can correct upto 16 bits of error, as it is applied on 8 decoder blocks in parallel (2 bit error correction for each block). So the data rate achieved considering only the data field (D) in this mode is: \\
$ 40 MHz \times 52 bits = 2.08 Gbps$ \\
In frame format without FEC, where error correction code is not used, the frame can carry $(52+64=116)$ bits of data out of 120 bits frame.
So in this mode data rate is measured: \\
$40 MHz \times 116 bits = 4.64 Gbps$ \\
Hence, the data transfer efficiency for the above mention two modes are $(2.08/4.80)\times 100 = 43.33\%$ and $(4.64/4.80 = 96.6\% $ respectively. 

Table~\ref{table:comparison} shows the comparison between our work and that of the other existing works. Existing system\cite{kadric:socc:2012} gives better speed with respect to our implementation but it does not incorporate any error correction mechanism as that of ours. 
\begin{table}[!t]
\begin{center}
\caption{Comparison with existing works }
\scalebox{.9}{

\begin{tabular}{|c|c|c|c|}
\hline 
\textbf{Device} & \textbf{Speed} & \textbf{Error correcting}  & \textbf{Line coding} \\ 
 \textbf{Used} & \textbf{(Gbps)} &\textbf{capability}& \textbf{used } \\
\hline 
Lattice SCM40\cite{Minami:ieeetran:2011} & 1.60 & Not mentioned & $8b/10b$ \\ 
\hline 
Altera Stratix IV\cite{kadric:socc:2012} & 8.50 & Not mentioned &  $8b/10b$ \\ 
\hline 
Altera EP2SGX90EF1152C5\cite{Liu:i2mtc:2013:X} & 2.125 & Not mentioned &  $8b/10b$ \\ 
\hline 
Virtex-II Pro series FPGA\cite{haoxu:nss:2007:XX} & 1.75 & CRC used for  &  $8b/10b$ \\ 
&&error detection only & \\
\hline 
\textbf{Kintex-7} &\textbf{ 4.80} & \textbf{Multi bit error correction}  & \textbf{Scrambler} \\ 
(\textbf{our design}) &&\textbf{BCH code used} & \\
\hline 
\end{tabular}}
%\caption{Description of the signals used in timing diagram}
\label{table:comparison}
\end{center}
\vspace{-20 pt}
\end{table}

\begin{figure}[htb]
%\hspace{-30 pt}
\centering
%\vspace{-10 pt}
\includegraphics[scale=0.35]{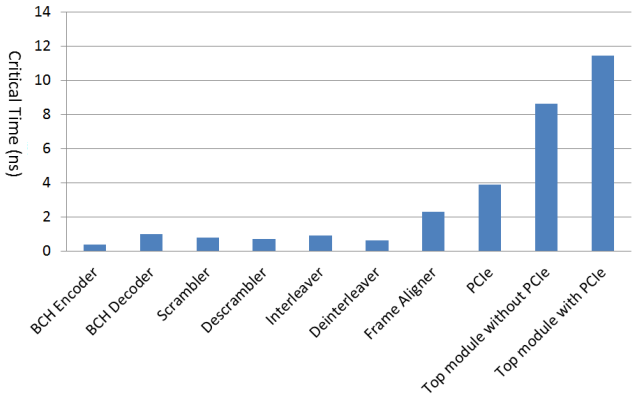}
\vspace{-45 pt}
\caption{Critical time of different blocks }
\label{fig:CiticalPathDelay}
\vspace{-15 pt}

\end{figure}

\begin{table}[t!]
\caption{Resource Utilization}
\scalebox{0.9}{
\begin{tabular}{|c|c|c|c|c|c|}
\hline 
\textbf{Board} & \textbf{Module} & \textbf{Slice} & \textbf{Slice} &\textbf{ LUT} & \textbf{BRAM} \\ 
&\textbf{Name} &\textbf{Register} &\textbf{LUTs} & \textbf{Flip Flop}& \\
\hline 
\multirow{14}{*}{\rotatebox{90}{\textbf{Kintex 7- 325t}}} & BCH & 7//407600 & 951/203800& 0 &7/951 \\ 
& Encoder (15,7,2) &&&& \\
\cline{2-6}
 & BCH & 135/407600 & 446/203800 & 0 & 119/462 \\
& Decoder (15,7,2) &&&&(25\%) \\
 \cline{2-6}
 & Scrambler & 52 & 53 & 5 & 0 \\
 \cline{2-6}
  &Descrambler &104 & 56 & 5 & 0 \\
 \cline{2-6}
&Interleaver &44 & 40 & 40 & 0 \\
 \cline{2-6}
&DeInterleaver &201 & 82 & 80 & 0 \\
 \cline{2-6}
&Frame Aligner &115& 308 & 72 & 0 \\
 \cline{2-6}
 &Encryption (AES) &1311& 4300 & 864 & 0 \\
 \cline{2-6}
  &Encryption (RSA) &116	& 31612 & 75 & 0 \\
 \cline{2-6}
 &PCIe &5882 & 5287 & 2694 & 10 \\
 \cline{2-6}
 &Top Module&3665 &9003 & 1998 & 5 \\
  & Without PCIe & & &  &\\
 \cline{2-6}
 &Top Module&8360 & 8555 & 3779 & 26 \\
 & With PCIe & & & & \\
 \cline{2-6}  
 \hline
\end{tabular} }
%\caption{Resource Utilization}
\vspace{-15pt}
\label{table:resource}
\end{table}  
Resource utilization for each functional block of the proposed DAQ system is given in Table~\ref{table:resource}. In Figure~\ref{fig:CiticalPathDelay} we show the critical time, which is the maximum delay time to get the output of a circuit for each of the circuit blocks. Power consumption is estimated using Xilinx Xpower tool and we show the estimated average logic and signal power for the various model of the proposed design in Table~\ref{table:powerconsumtion}. To the best of our knowledge, we are reporting the critical time and power consumption of this type of system for the first time.
\par
\begin{figure}[t!]
\centering
\includegraphics[scale=0.3]{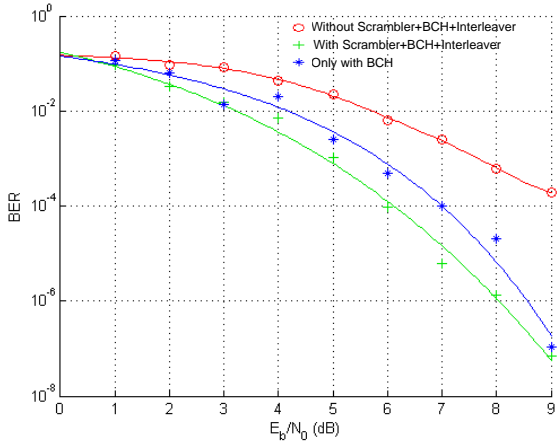}
\vspace{-10 pt}
\caption{Study of of BER with noise}
\vspace{-18 pt}
\label{Rxmarginanalysis}
\end{figure}

 SEU error is random in nature. We have emulated the SEU error by generating random error on the input data using random error generator~\cite{Antoni:IEEETran:2003}. The simulation results of BER is shown in the Figure~\ref{Rxmarginanalysis} with respect to the noise (Eb/N), which varies from 0 dB to 10 dB. We have assumed the power spectral density of noise as a Poison distribution. Figure~\ref{Rxmarginanalysis} shows the efficiency of our system comprising of BCH code with interleaver and scrambler gives the best performance in presence of noise. The throughput of the DAQ system is measured as 4.8 Gbps in the Xilinx platform installed in Fedora OS.
%===============================
\begin{table}[htb]
\centering
\caption{Module wise power consumption}
\scalebox{1.2}{
\begin{tabular}{|c|c|c|c|}
\hline 
\textbf{Board} & \textbf{Module } & \textbf{Logic } &\textbf{Signal}\\ 

&\textbf{Name} &\textbf{Power(mW)} & \textbf{Power(mW)}\\
\hline 
\multirow{10}{*}{\rotatebox{90}{\textbf{Kintex 7-325t}}} & BCH  & 0.02 & 0.01\\
& Encoder(15,7,2)&& \\
 \cline{2-4}
 & BCH  & 0.05 & 0.07 \\
&Decoder(15,7,2)&& \\
 \cline{2-4}
 & Scrambler & 0.04 & 0.00 \\
 \cline{2-4}
  &Descrambler &0.01 & 0.00 \\
 \cline{2-4}
&Interleaver &0.01 & 0.01\\
 \cline{2-4}
&DeInterleaver & 0.01& 0.02 \\
 \cline{2-4}
&Frame Aligner &1.34& 1.07 \\
 \cline{2-4}
 &PCIe &253.24 & 45.55 \\
 \cline{2-4}
 &Top Module& 474.18 & 2.91 \\
  & Without PCIe & &  \\
 \cline{2-4}
 &Top Module& 304.24 & 56.31 \\
 & With PCIe & &  \\
 \cline{2-4}  
 \hline
\end{tabular} }
%\vspace{-5 pt}
%\caption{Module wise power consumption}
\label{table:powerconsumtion}
\vspace{-20 pt}
\end{table}   
%===============================
%=================================================================
\section{Conclusion}\label{Conclusion}
In this work we have proposed a novel DAQ design for HEP experiments. The proposed DAQ supports high speed (in terms of Gbps) optical data communication and also corrects multi-bit error. The DAQ design has been implemented on Xilinx Kintex-7 board and real test setup has been developed involving board to board communication and PCIe interfacing with a host PC. A detailed performance analysis of the DAQ implementation is presented in terms of timing diagram, resource utilization and critical timing for of each of the blocks (FPGA), power consumption and BER. The proposed DAQ design and its implementation involving optical data communication and multi-bit error correction capability can be considered as first of its kind and can serve as a benchmark design in HEP DAQ.  In future, we plan to use the concept of multi-pocessor into this system for more efficiency in data processing and test the setup in the radiation zone. 
\section*{Acknowledgment}
The authors like to acknowledge  DAE, DST, GSI, CERN, TEQIP-II (CU) to provide necessary support for carrying out the research.
%\section{Conclusions}

%ACKNOWLEDGMENTS are optional
%\section{Acknowledgments}

%\bibliographystyle{unsrt}
\bibliographystyle{unsrt}
\bibliography{IEEEexample}

\begin{thebibliography}{10}

\bibitem{Wang:iciea:2009}
Wang Lixin, Song Wei, and Lv~Chao.
\newblock Implementation of high speed real time data acquisition and transfer
  system.
\newblock In {\em Industrial Electronics and Applications, 2009. ICIEA 2009.
  4th IEEE Conference on}, pages 382--386, May 2009.

\bibitem{Minami:ieeetran:2011}
S.~Minami, J.~Hoffmann, N.~Kurz, and W.~Ott.
\newblock Design and implementation of a data transfer protocol via optical
  fiber.
\newblock {\em Nuclear Science, IEEE Transactions on}, 58(4):1816--1819, Aug
  2011.

\bibitem{kadric:socc:2012}
E.~Kadric, N.~Manjikian, and Z.~Zilic.
\newblock An fpga implementation for a high-speed optical link with a pcie
  interface.
\newblock In {\em SOC Conference (SOCC), 2012 IEEE International}, pages
  83--87, Sept 2012.

\bibitem{Liu:i2mtc:2013:X}
Liansheng Liu, Chuan Liu, Yu~Peng, and Datong Liu.
\newblock A design of fibre channel node with pci interface.
\newblock In {\em Instrumentation and Measurement Technology Conference
  (I2MTC), 2013 IEEE International}, pages 1817--1822, May 2013.

\bibitem{haoxu:nss:2007:XX}
Hao Xu, Zhan'an Liu, Yunpeng Lu, Lu~Li, Dixin Zhao, and Ya'nan Guo.
\newblock Fpga based high speed data transmission with optical fiber in trigger
  system of besiii.
\newblock In {\em Nuclear Science Symposium Conference Record, 2007. NSS '07.
  IEEE}, volume~1, pages 818--821, Oct 2007.

\bibitem{mattihalli:cecnet:2012}
C.~Mattihalli.
\newblock Design and realization of serial front panel data port (sfpdp)
  protocol.
\newblock In {\em Consumer Electronics, Communications and Networks (CECNet),
  2012 2nd International Conference on}, pages 2505--2509, April 2012.

\bibitem{bohm:nss:mic:2012}
H.~Kavianipour and C.~Bohm.
\newblock High performance fpga-based scatter/gather dma interface for pcie.
\newblock In {\em Nuclear Science Symposium and Medical Imaging Conference
  (NSS/MIC), 2012 IEEE}, pages 1517--1520, Oct 2012.

\bibitem{Siewiorek:CED}
Daniel~P. Siewiorek and Robert~S. Swarz.
\newblock {\em Reliable Computer Systems (3rd Ed.): Design and Evaluation}.
\newblock A. K. Peters, Ltd., Natick, MA, USA, 1998.

\bibitem{TMR:IeeeTran:NuclrPhy}
A~Gabrielli, G.~De~Robertis, D.~Fiore, F.~Loddo, and A~Ranieri.
\newblock Architecture of a slow-control asic for future high-energy physics
  experiments at slhc.
\newblock {\em Nuclear Science, IEEE Transactions on}, 56(3):1163--1167, June
  2009.

\bibitem{chien:search}
R.T. Chien.
\newblock Cyclic decoding procedures for bose- chaudhuri-hocquenghem codes.
\newblock {\em Information Theory, IEEE Transactions on}, 10(4):357--363, 1964.

\bibitem{Varghese:multibit:error}
B.~Varghese, S.~Sreelal, P.~Vinod, and AR. Krishnan.
\newblock Multiple bit error correction for high data rate aerospace
  applications.
\newblock In {\em Information Communication Technologies (ICT), 2013 IEEE
  Conference on}, pages 1086--1090, April 2013.

\bibitem{Antoni:IEEETran:2003}
L.~Antoni, R.~Leveugle, and B.~Feher.
\newblock Using run-time reconfiguration for fault injection applications.
\newblock {\em Instrumentation and Measurement, IEEE Transactions on},
  52(5):1468--1473, Oct 2003.

\end{thebibliography}

\end{document}